\documentclass[reprint, amsmath, amssymb, aps, showkeys]{revtex4-2}

\usepackage{gensymb}
\usepackage{booktabs}
\usepackage{graphicx}
\graphicspath{{./Figures/}}
\usepackage{multirow}
\usepackage{dcolumn}
\usepackage{bm}
\usepackage{times}
\usepackage{algorithm}
\usepackage{algorithmicx}
\usepackage{algpseudocode}
\usepackage{listings}
\usepackage{natbib}
\usepackage{placeins}
\usepackage{longtable}
\usepackage{array}
\usepackage[explicit]{titlesec}

\usepackage[colorlinks = true,
linkcolor = blue,
urlcolor  = blue,
citecolor = blue,
anchorcolor = blue]{hyperref}

\begin{document}

\preprint{APS/123-QED}

\title{Large Language Models for Material Property Predictions:\\ elastic constant tensor prediction and materials design}

\author{Siyu Liu$^{1,2}$}
\author{Tongqi Wen$^{1,2}$}\email{tongqwen@hku.hk}
\author{Beilin Ye$^{1}$}
\author{Zhuoyuan Li$^{1,2}$}
\author{David J. Srolovitz$^{1,2}$}\email{srol@hku.hk}

\affiliation{$^{1}$Center for Structural Materials, Department of Mechanical Engineering, The University of Hong Kong, Hong Kong SAR, China}
\affiliation{$^{2}$Materials Innovation Institute for Life Sciences and Energy (MILES), The University of Hong Kong, Shenzhen, China}

\date{\today}

\begin{abstract}

Efficient and accurate prediction of material properties is critical for advancing materials design and applications. 
The rapid-evolution of large language models (LLMs) presents a new opportunity for material property predictions, complementing experimental measurements and multi-scale computational methods. 
We focus on predicting the elastic constant tensor, as a case study, and develop domain-specific LLMs for predicting elastic constants and for materials discovery. The proposed ElaTBot LLM enables simultaneous prediction of elastic constant tensors, bulk modulus at finite temperatures, and the generation of new materials with targeted properties. 
Moreover, the capabilities of ElaTBot are further enhanced by integrating with general LLMs (GPT-4o) and Retrieval-Augmented Generation (RAG) for prediction. 
A specialized variant, ElaTBot-DFT, designed for 0 K elastic constant tensor prediction, reduces the prediction errors by 33.1\% compared with domain-specific,  material science LLMs (Darwin) trained on the same dataset. 
This natural language-based approach lowers the barriers to computational materials science and highlights the broader potential of LLMs for material property predictions and inverse design.

\end{abstract}

\maketitle

Property data are essential for determining the suitability of materials for specific applications. 
For example, flexible electronics require materials with targeted elastic stiffness~\cite{elasticity1}, thermal management systems rely on materials with sufficiently high thermal conductivity~\cite{thermal1}, and electronic devices depend on materials with appropriate band structure~\cite{bandgap1}. 
Given the diverse property profile required for individual applications, deep understanding or at least robust property prediction would be a great aid to material selection and/or alloy design. 
While the former is a long-term goal of materials science, high-throughput material property measurements and prediction, coupled with new techniques in artificial intelligence (AI) have the potential to revolutionize materials development 
in the short term.

While experimental approaches for determining materials properties remains the gold standard, they are often hindered by expense and the time required to synthesize materials and measure properties (and, at times, lead to results that are either inconsistent or not sufficiently accurate), such as in the case of elastic constant measurements~\cite{WU202041}. 
Recent advancements in simulation techniques and computational power have made computational modeling a critical tool for property prediction.
Given the diversity of length and time scales that control material properties, multi-scale modeling has emerged as an often efficient and sufficiently accurate approach for  materials property prediction. 
For example, atomistic simulations with quantum-mechanical accuracy can accurately predict the full elastic constant tensors and/or band gaps (using hybrid exchange-correlation functionals), while phase-field modeling and other continuum based-methods enable microstructure evolution and defect property prediction. 
However, challenges (e.g., data transfer and error propagation) often remain significant obstacles to achieving accurate, macroscopic  predictions in multi-scale modeling frameworks. 
The emergence of large language models (LLMs) presents a new opportunity for materials property prediction, with the potential to bridge  gaps between experiment data (e.g., sourced from literature databases) and computational materials simulation approaches~\cite{fan_2024_as}.

LLMs, for example ChatGPT, have demonstrated some remarkable successes across a wide range of materials applications, including high-throughput discovery of physical laws~\cite{physical-law}, generation of metal-organic frameworks (MOFs)~\cite{mofgpt}, design of chemical reaction workflows~\cite{chemcrow}, determining crystal structure (CIF, crystallographic information file)~\cite{cif}, electron microscopy image analysis~\cite{electronic-microscope}, and guiding automated experiments~\cite{crest}. 
LLMs achieve this by leveraging their capabilities such as rapid literature summarization~\cite{literature}, prompt engineering~\cite{LIU2024} and/or integration with external tools~\cite{toolagent}. 
This approach can make them  superior to traditional machine learning (ML) models, particularly when dealing with complex and multitask processes at scale~\cite{zhang2024scientific}. 
One major strength of LLMs is their foundation in natural language-based training, fine-tuning, and application, which lowers the barrier to entry for researchers without a strong background in computer science or coding~\cite{ouyang2024}. 
Moreover,  underlying pretrained models encode extensive materials science knowledge, giving LLMs remarkable ability in cases where datasets are sparse  (through transfer learning)—an achievement that previously required highly specialized algorithms~\cite{game-changer}.

\begin{figure*}[htbp]
	\centering
	\includegraphics[width=0.95\textwidth]{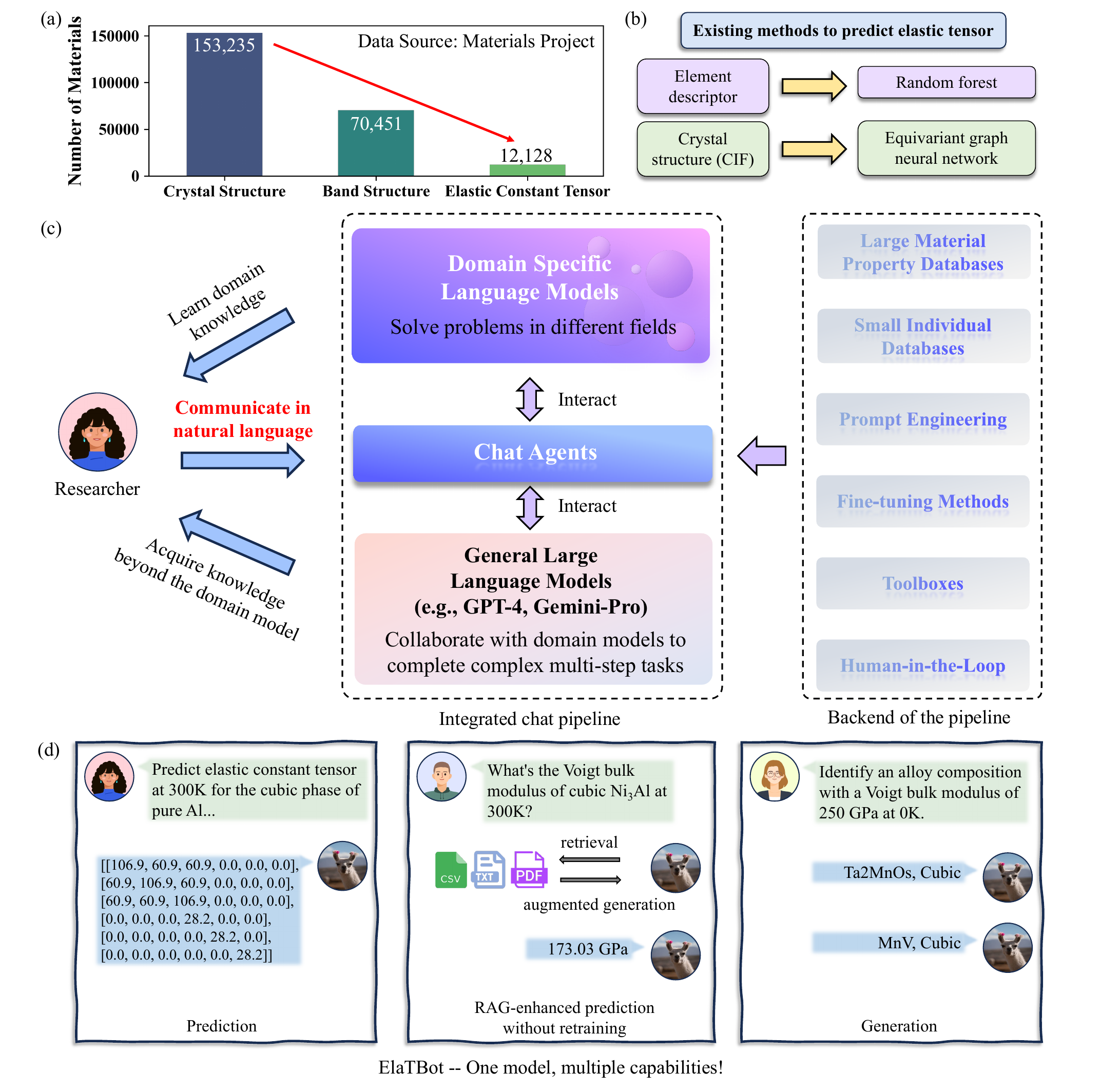}
	\caption{Datasets and overview of the ElaTBot for predicting elastic properties. (a) Comparison of the number of materials in the Materials Project database~\cite{mp_2013_aplm} with available data on crystal structures, band structures, and elastic properties. The availability of elastic constant tensors data is significantly lower than that of crystal and band structure data. (b) Overview of existing methods used to predict elastic constant tensors, which primarily relied on element descriptors and structural features (constructed by CIF). (c) A flowchart illustrating the process of using large language models (LLMs) to acquire material knowledge. This method enables researchers to gain domain-specific insights, allowing those without extensive programming skills or theoretical expertise to conduct research, thereby lowering the entry barrier into materials science. (d) Capabilities of our specialized LLM ElaTBot. By incorporating elastic constant tensors data at finite temperatures, we develop an LLM-based agent, ElaTBot, which is capable of predicting elastic constant tensors, enhancing prediction without retraining by leveraging external tools and datasets, and generating chemical composition for materials with specific modulus.
}\label{fig1}
\end{figure*}

Despite these  advances in LLM materials science applications, LLM performance in predicting quantitative material properties has been limited, especially when faced with small datasets~\cite{jacobs2024regressionlargelanguagemodels}. 
For example, Jablonka et al.~\cite{Jablonka2024} showed that while LLMs can  predict properties like HOMO-LUMO gaps, solubility, photoswitching behavior, solvation free energies, and photoconversion efficiency,  the results were no better than with traditional ML models. 
Enhancing the quantitative prediction capabilities of LLMs, while leveraging their strengths in natural language interaction and multitasking, can significantly expand their potential in materials science applications.

In this work, we focus on predicting the elastic constant tensor as a case study of quantitative prediction of a material property. 
The elastic constant tensor is a fundamental property that describes the elastic response of materials to external forces~\cite{anie.202110716}
and serves as a indicator of the nature of  intrinsic bonding within a material~\cite{LEVINE20101530}. 
Mechanical (Young’s modulus, Poisson’s ratio,...), thermal (thermal conductivity), and acoustic (sound velocity) properties can all be derived starting from the elastic constant tensor~\cite{D3DD00233K} (often together with other basic material properties). 
Here, we introduce ElaTBot and ElaTBot-DFT (DFT is quantum mechanical density functional theory), LLMs developed through prompt engineering and knowledge fusion training. 
ElaTBot is designed to predict elastic constant tensors, bulk modulus at finite temperatures, and propose  materials with specific elastic properties. 
To our knowledge, ElaTBot is the first model capable of directly and efficiently predicting the full elastic constant tensor at finite temperatures. ElaTBot-DFT, a variant specialized for 0 K elastic constant tensor prediction, reduces prediction error by 33.1\% compared to the material science LLM Darwin~\cite{xie2023darwin} using the same training and test sets. 
These results highlight the potential of LLMs for numerical materials property predictions.

\begin{figure*}[htbp]
  \centering
  \includegraphics[width=0.95\textwidth]{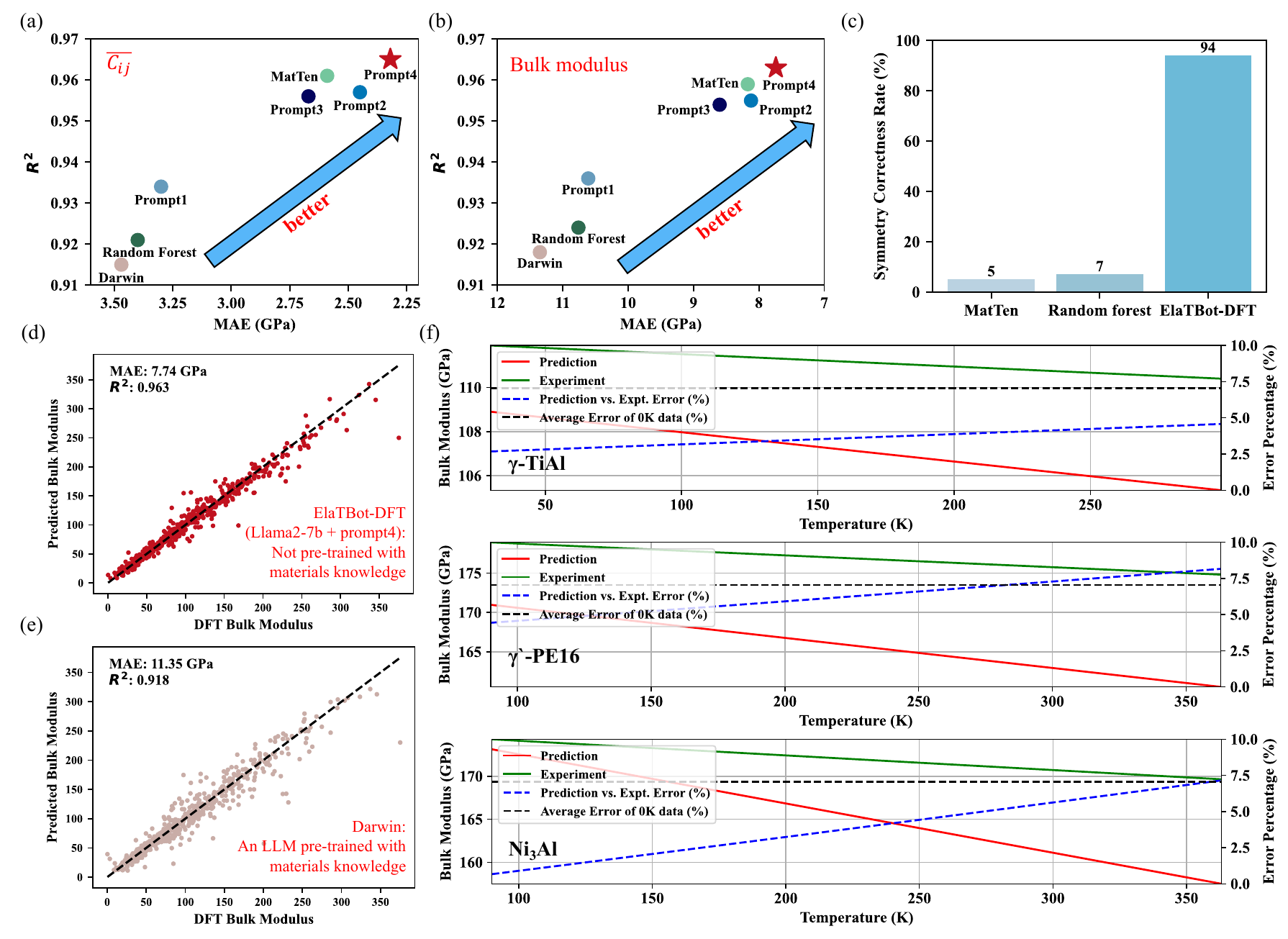}
  \caption{Prediction abilities of ElaTBot-DFT and ElaTBot. (a, b) Performance comparison of the Llama2-7b model using different prompt types, the MatTen model, random forest model, and Darwin model in predicting $\overline{C_{ij}}$ (GPa) and bulk modulus based on MAE and $R^2$ on the 0K DFT test set, all trained on the same dataset. (c) Symmetry validation for elastic constant tensors predicted by MatTen, random forest, and ElaTBot-DFT models. Symmetry correctness is defined as components within $\pm$2 GPa where zero values are required by the Voigt format matrix. (d, e) Performance comparison of ElaTBot-DFT model against the pre-trained Llama2-7b model (Darwin, trained with a materials knowledge database) for bulk modulus prediction, using the same test set and training data with prompt type 4. (f) The capability of ElaTBot to predict finite temperature bulk modulus. The red line indicates predicted values, the green line shows  experimental data~\cite{NiAl, gamma-TiAl}, the blue dashed line indicates the percentage error trend, and the black dashed line shows the average error (7.05 \%) for the 0K temperature test set.
}\label{fig2}
\end{figure*} 

\section*{Training Specialized LLMs}

Despite the importance of elastic constant tensors in materials science, complete elastic constant tensor data for inorganic crystals remains scarce due to experimental and computational limitations. 
Fig.~\ref{fig1}(a) shows that elastic constant data is scarce in the Materials Project~\cite{mp_2013_aplm};  $\sim$7.9\% as abundant as crystal structure data and $\sim$17.2\% as abundant as band structure data. 
The elastic constants are a fourth-rank tensor with as many as 21 independent components (in static equilibrium) which is often  represented as a symmetric $6\times6$ Voigt matrix $C_{ij}$~\cite{LI2022108280}. 
Predicting these components is far more complex than predicting scalar properties, such as formation energy, free energy, or bulk modulus. 
Fig.~\ref{fig1}(b) lists a few ML approaches for predicting elastic constant tensors. 
Chemical composition-based upon element descriptors were used to predict specific components or $C_{ij}$~\cite{REVI2021110671, VAZQUEZ2022117924}, but  not the full elastic constant tensor. 
More recent models that leverage crystal structure descriptors to predict the full elastic constant tensor face challenges such as long training times and complex model architectures~\cite{D3DD00233K, PhysRevResearch.5.043198}. 
These models are often restricted to single-task predictions of elastic properties and lack the capabilities to propose new materials tailored to specific properties or learn from new data without retraining.

Fig.~\ref{fig1}(c) presents an integrated approach, combining ML and natural language processing, for predicting material properties and identifying materials with targeted properties. 
Specifically, for elastic properties prediction and materials generation, we developed two domain-specific LLMs: ElaTBot and ElaTBot-DFT, which predict elastic properties such as the elastic constant tensor, bulk, shear and Young's moduli, as well as the Poisson ratio. 
To further improve user interaction and task handling, we implemented an AI-driven agent capable of utilizing tools and databases, and general LLMs to perform complex, multi-step tasks.
This agent can process new (and unseen) data by integration of external tools and vector databases. 
Its responses can be fed into general LLMs (e.g., GPT-4, Gemini) to further extend its capabilities and tackle more complex, multi-step tasks. 
Fig.~\ref{fig1}(d) shows three capabilities of our specialized LLM ElaTBot: prediction, Retrieval-Augmented Generation (RAG)-enhanced prediction
~\cite{RAG} without retraining, and generation. 

To train the ElaTBot, we first employ prompt engineering to translate structural and compositional information into textual descriptions, fine-tuning the general LLM Llama2-7b to create ElaTBot-DFT, a specialized model for DFT prediction of  elastic constant tensors at 0 K. 
ElaTBot-DFT serves as a benchmark for elastic constant tensor prediction, particularly since other models are limited in addressing finite-temperature predictions. 
Next, we employ several steps to  enhance LLM performance~\cite{gruver2024finetunedlanguagemodelsgenerate}, incorporating finite-temperature data and fusing this knowledge to develop ElaTBot. 
Prompt engineering leads to a reduction of the prediction error of the average value of the elastic constant tensor $\overline{C_{ij}}$  (see \emph{Methods}) by 33.1\% for ElaTBot-DFT compared to Darwin~\cite{xie2023darwin}, a materials science LLM built on the same dataset. 
Through knowledge fusion, ElaTBot accurately fits the temperature-dependent bulk modulus curves (derived from the elastic constant tensor) for new multicomponent alloys, with errors near room temperature approaching the average error for the 0K test set. 
RAG-enhanced~\cite{RAG} predictions with limited finite-temperature data further improves ElaTBot errors for the bulk modulus from 27.49\% to 0.95\% across nine alloys at various temperatures without retraining or fine-tuning.

We  integrate ElaTBot with GPT-4o to propose/screen materials based upon bulk modulus and other requirements of targeted applications. 
These include materials with low corrosion rates and high biocompatibility (measured by median lethal dose) with a bulk modulus similar to that of bone for bone implantation, high bulk but low shear modulus materials suitable for  exoskeletons of soft robots, corrosion-resistant materials suitable for saline environments, and materials for ion-conductive protective capping layers.

\section*{Elastic Constant Tensor Predictions
}
Our model uses text-based inputs to predict the elastic constant tensor; this  requires carefully designed prompt templates. 
We develop a  prompt (prompt type 4 in Supplementary Table~S1) that incorporates both chemical composition and crystal structure descriptions as inputs to the model (Llama2-7b, an open-source general LLM). 
These textual descriptions were generated by extracting relevant data from the material composition and structure and converting these into natural language. 

We conducted a series of 
``experiments'' to assess the effects of different input formats on model performance: JSON-formatted (JavaScript Object Notation) structure descriptions (prompt type 1), textual descriptions of crystal structure (prompt type 2),  textual descriptions of the composition  (prompt type 3), and textual description of both chemical composition and crystal structure (prompt type 4). 
The model was trained using a 0K density functional theory (DFT) dataset containing 9,498 materials with elastic constant tensor data from the Materials Project, with 500 materials for validation and 522 for testing. 
Fig.~\ref{fig2}(a) and Supplementary Table~S2 show that prompt type 4 achieves a mean absolute error (MAE; the average of the absolute differences between the predicted and actual values for all data points) of 2.32 GPa and $R^2$ of 0.965 for predicting the average elastic constant tensor component ($\overline{C_{ij}}$), outperforming other prompt types (explicit definitions of the  MAE and $\overline{C_{ij}}$ are in  \emph{Methods}).
Compared to prompt type 1 (JSON format), prompt type 4 reduces the MAE by 29.7\% and increases $R^2$ by 3.3\%. 
When compared to prompt type 2 (crystal structure descriptions only), prompt type 4 achieves a 5.3\% reduction in MAE and a 0.8\% increase in $R^2$. 
Compared to prompt type 3 (composition descriptions only), prompt type 4 yields a 13.1\% reduction in MAE and a 0.9\% increase in $R^2$. 
These results demonstrate that LLMs perform better when trained with natural language-like inputs, and that using both structural and compositional information improves elastic constant tensor prediction. 
The  bulk modulus results (derived from the elastic constant tensor) in Fig.~\ref{fig2}(b)  confirm this: prompt type 4 achieves an MAE of 7.74 GPa and an $R^2$ of 0.963, representing a 27\% reduction in MAE and a 2.9\% increase in $R^2$ compared to prompt type 1. 
Therefore, prompt type 4 was selected for training Llama2-7b for our ElaTBot-DFT model. 

  \begin{figure*}[htbp]
    \centering
    \includegraphics[width=0.95\textwidth]{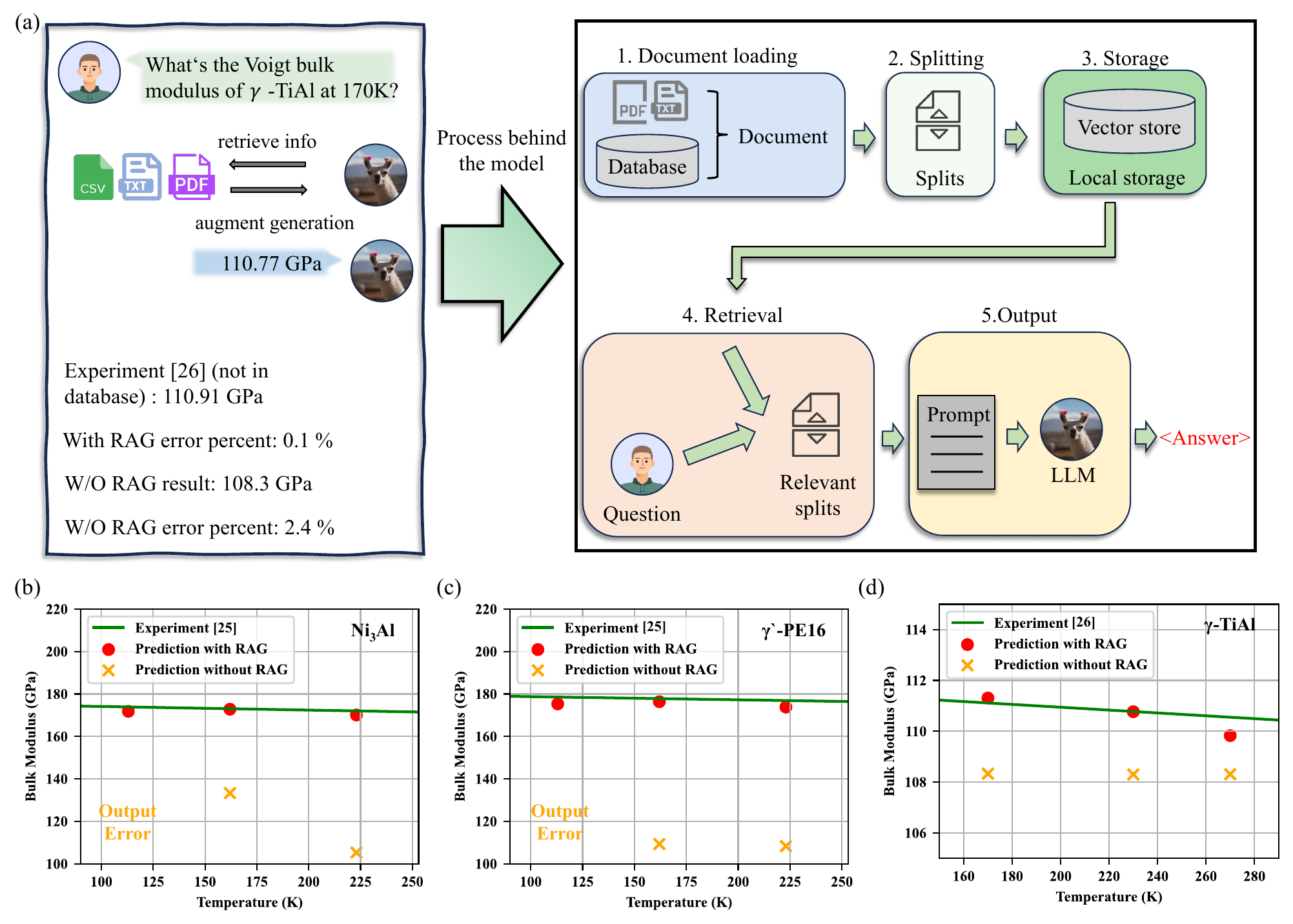}
    \caption{Integration of Retrieval-Augmented Generation (RAG) with ElaTBot for enhanced prediction. (a) The steps involved in enabling ElaTBot to perform RAG are as follows: 1. Document loading: external documents or data sources are ingested into the system for further use; 2. Splitting: the documents are broken down into smaller, manageable chunks to optimize retrieval; 3. Storage: these chunks are stored in an indexed format for fast and efficient searching; 4. Retrieval: the system identifies and retrieves the most relevant chunks in response to the the query; 5. Output: ElaTBot generates a more accurate and informed response by incorporating the retrieved information. (b, c, d) Differences in predicted bulk modulus with and without the RAG method for Ni$_3$Al, $\gamma^{'}-PE16$ (Ni$_{72.1}$Al$_{10.4}$Fe$_{3.2}$Cr$_{1.0}$Ti$_{13.3}$) and $\gamma-TiAl$ (Ti$_{44}$Al$_{56}$) as a function of temperature. The bulk modulus error percent decreased from 27.49\% to 0.95\% in 9 alloys with different temperatures after using RAG.}\label{fig3}
  \end{figure*}

We compared the performance of ElaTBot-DFT with two widely-used models for predicting the full elastic constant tensor: the random forest model, which utilizes Magpie (Materials Agnostic Platform for Informatics and Exploration) features based on composition~\cite{Ward2016}, and the MatTen model, which employs a crystal structure graph neural network~\cite{D3DD00233K}. 
As shown in Figs.~\ref{fig2}(a, b), Supplementary Fig.~S1 and Table~S3, when trained on the dataset, ElaTBot-DFT using prompt type 4 achieves a 31.8\% reduction in MAE and a 4.8\% increase in $R^2$ for predicting the average elastic constant tensor components ($\overline{C_{ij}}$) compared to the random forest model. 
Compared to the MatTen model, ElaTBot-DFT reduces MAE by 10.4\% and improves $R^2$ by 0.4\%. 
This demonstrates that, even with a relatively small dataset, LLMs trained with well-designed textual descriptions can outperform traditional methods, contrary to previous studies using QA-based training approaches~\cite{Jablonka2024}. 
We also examined the symmetry of the generated elastic constant tensors that result from the rigorous application of crystal symmetries; this symmetry requires certain $C_{ij}$ components to be zero and a fixed relationship between some others. 
Under strict criteria (error margin of $\pm$2 GPa), ElaTBot-DFT achieves a symmetry accuracy of 94\%, significantly outperforming MatTen (5\%) and the random forest model (7\%) (Fig.~\ref{fig2}(c)).
Traditional numerical models tend to produce small non-zero values due to algorithmic limitations, while the natural language-based model, ElaTBot-DFT, accurately outputs a ``0'' where appropriate.

  We further compared ElaTBot-DFT predictions with those from the domain-specific materials LLM, Darwin, for elastic constant tensor prediction. 
Domain-specific LLMs are widely believed to outperform general LLMs on specialized problems~\cite{xie2023large}; however, as shown in Figs.~\ref{fig2}(a, b, d, e) and Table~S3, Darwin (even after fine-tuning on the same dataset) underperforms  ElaTBot-DFT in predicting the $\overline{C_{ij}}$ and bulk modulus. 
Specifically, the MAEs of ElaTBot-DFT are 33.1\% and 31.8\% lower than those of Darwin for $\overline{C_{ij}}$ and bulk modulus, respectively. 
This suggests that integrating the reasoning abilities of a general LLM with fine-tuning on a specific dataset may yield better results for tasks requiring quantitative property predictions. 
Fine-tuning a model with domain-specific knowledge (like Darwin) can lead to gaps in its abilities and knowledge loss, which may reduce the effectiveness in specialized tasks~\cite{li2024examiningforgettingcontinualpretraining}.

  \begin{figure*}[htbp]
    \centering
    \includegraphics[width=0.95\textwidth]{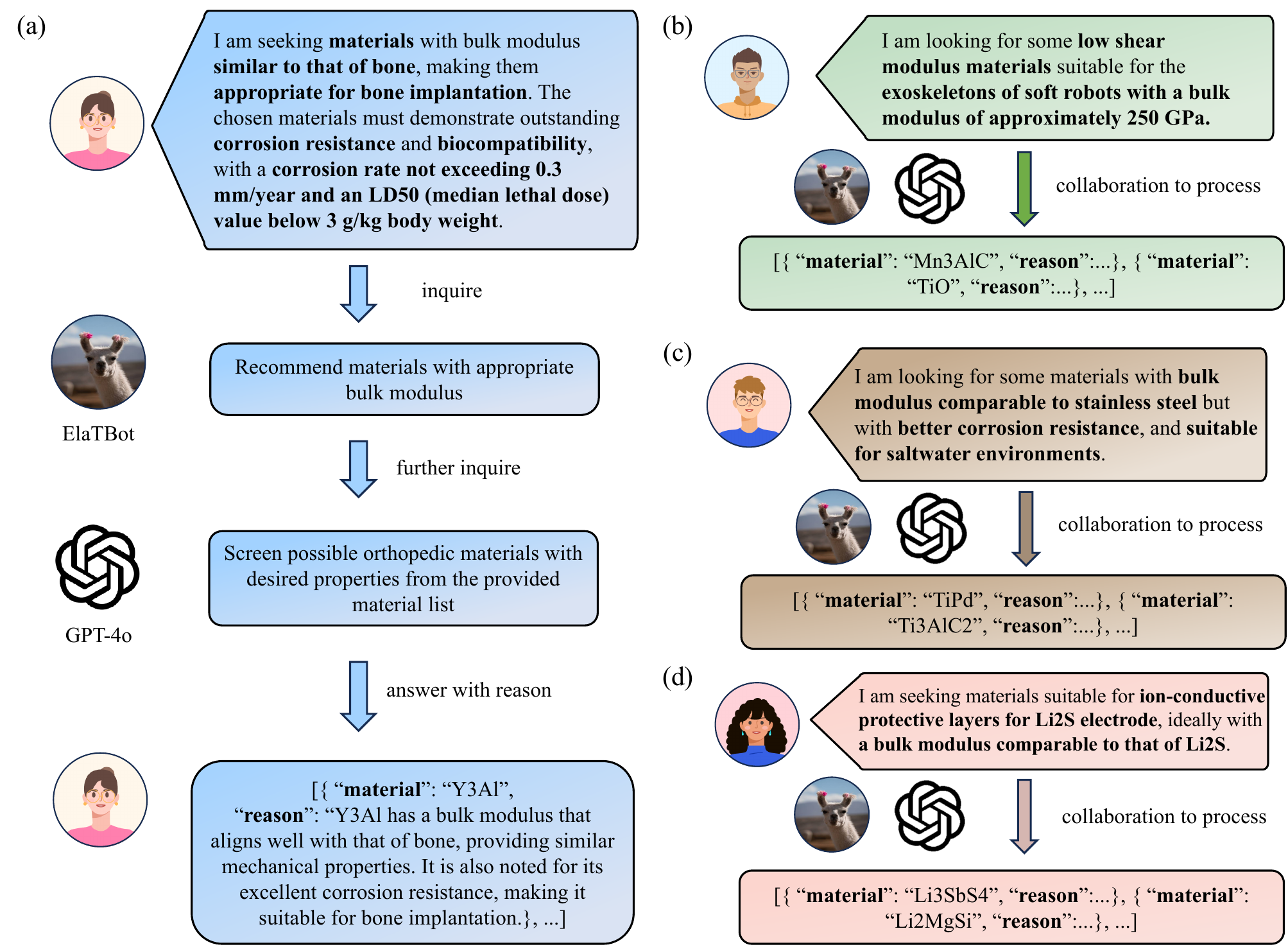}
    \caption{Integration of a domain-specific LLM (ElaTBot) with a general LLM (GPT-4o) for materials discovery. 
 The process begins by requesting ElaTBot to generate material compositions with a target bulk modulus. 
 Next, GPT-4o refines the results to identify compositions that meet specific application requirements. 
 Examples of applications include: (a) generating orthopedic materials with bulk modulus similar to bone, (b) discovering materials with high bulk modulus and low shear modulus, (c) finding new corrosion-resistant materials, and (d) identifying materials for ion-conductive protective layers.}\label{fig4}
  \end{figure*}

  Finally, we integrated the finite-temperature dataset and designed four tasks (Supplementary Table~S4) with corresponding training text inputs (elastic constant tensor prediction, bulk modulus prediction, material generation based on bulk modulus, and text infilling) to conduct multi-task knowledge fusion training. 
This approach equips ElaTBot with multiple capabilities, including the ability to predict elastic constant tensors at finite temperatures. 
Although the text infilling task does not directly predict material properties, previous studies have shown that it improves the overall multi-task performance~\cite{gruver2024finetunedlanguagemodelsgenerate}. 
To test the effectiveness of ElaTBot, we selected three  multicomponent alloys not in the training set (cubic  $\text{Ni}_{3}\text{Al}$, $\gamma'-\text{PE16}$ (Ni$_{72.1}$Al$_{10.4}$Fe$_{3.2}$Cr$_{1.0}$Ti$_{13.3}$), and tetragonal  $\gamma-\text{TiAl}$ (Ti$_{44}$Al$_{56}$)) and predicted their bulk modulus as a function of temperature (based on the full elastic constant tensors). 
Given the limited finite-temperature training data--just 1,266 samples--and the vast compositional space of alloys, predicting accurate values over a wide range of temperature and compositions is inherently challenging. 
We predicted the bulk modulus at 11 temperatures for $\text{Ni}_{3}\text{Al}$ and $\gamma'-\text{PE16}$ and 15 temperatures for $\gamma-\text{TiAl}$.  
Fig.~\ref{fig2}(f) shows that the fitted lines from the predictions of ElaTBot closely align with experimental data~\cite{NiAl, gamma-TiAl}, particularly at low temperatures, where ElaTBot exhibits smaller errors. This demonstrates that incorporating the 0K DFT dataset from the Materials Project helps reduce prediction errors, highlighting the effectiveness of the multi-task knowledge fusion training approach.


\section*{RAG Enhanced Predictions
}

Retrieval-Augmented Generation (RAG)
~\cite{RAG} provides an effective method for LLMs to access up-to-date databases, enabling the most current RAG-enhanced prediction \emph{without}  model retraining. 
RAG integrates information retrieval with generative models, enhancing the knowledge scope and accuracy of LLM~\cite{RAG} output. 
The retrieval module extracts relevant data from external sources, which is then combined with the generative model to deliver more accurate predictions or text generation (Fig.~\ref{fig3}(a)). 
This approach allows LLMs to stay current with new data or literature, by incorporating knowledge rather than solely relying on  pre-trained models.

Fig.~\ref{fig3}(a) compares the ElaTBot bulk modulus predictions for $\gamma-TiAl$ at 170 K with and without RAG support. 
Since the finite temperature data was not in the original ElaTBot training set, the model automatically queries our external database, finds bulk modulus data for $\gamma-TiAl$ at similar temperatures (the 170 K data was removed from the database for comparison purposes). 
The predicted value (110.77 GPa) differs by only 0.1\%  from the true value, whereas without RAG, the error increases to 2.4\%. 
To ensure a fair comparison, we customized the RAG prompt in order to isolate  its influence from training prompts. 
Further testing on alloy data at various temperatures (see Figs.~\ref{fig3}(b, c, d) and Table~S5) demonstrates that RAG reduces the average error from 27.49\% to 0.95\%. 
By incorporating RAG, ElaTBot achieves RAG-enhanced prediction capabilities, allowing it to reason well beyond its training set that contained minimal similar data.

\section*{Material Discovery
}

The knowledge-fused ElaTBot may also be applied to  inverse materials design. 
By combining the domain-specific LLM (ElaTBot) with a general LLM (GPT-4o), we can search for materials with specific bulk modulus for various applications. Fig.~\ref{fig4}(a) shows an example where we identify orthopedic materials for bone replacement with a bulk modulus similar to that of bone ($<$50 GPa)~\cite{WU202041}. 
Using the ``material generation task'' prompt template from Table~S4, we identify the bulk modulus target as  $<$50 GPa and request the ElaTBot agent to generate potential material compositions. 
This process can be automated via a Python script to obtain multiple material compositions. 
The generated compositions are then passed to GPT-4o (dialogue record in Fig.~S3) with more specific requirements, such as corrosion resistance and biocompatibility; i.e., a corrosion rate $<0.3$ mm/year and an LD$_{50}$ (median lethal dose)  $<3$ g/kg body weight~\cite{Gorejova2019}. 
This results in a list of compositions that satisfy both the bulk modulus and orthopedic material criteria, along with explanations for each recommendation. 
Figs.~\ref{fig4}(b, c, d) and Figs.~S4-S6 show that this process can be extended to discover materials with high bulk modulus ($\sim$250 GPa) and low shear modulus, new corrosion-resistant materials with bulk modulus similar to stainless steel ($\sim$160 GPa)~\cite{engineeringtoolbox}, or materials for ion-conductive protective layers for Li$_{2}$S with a $\sim$40 GPa bulk modulus (similar to that of Li$_{2}$S~\cite{deng2015elastic}) that
could be used to stabilize high-capacity battery electrodes by providing structural support and accommodating volume changes during charge/discharge cycles in lithium-sulfur batteries. 
These applications demonstrate the potential of integrating domain-specific and general LLMs to accelerate new material discovery and design.

Based on the prediction and generation abilities above, we developed a multi-function agent interface (Fig.~S2) that allows researchers to predict, generate, or engage in RAG-enhanced prediction through natural language dialogue, without the heavy load of coding.

\section*{Discussion}

We demonstrated the potential of LLMs in predicting  elastic constant tensors and discovering materials with targeted elastic properties. 
By introducing ElaTBot-DFT (to accurately predict elastic constants at 0K) and ElaTBot (extention to finite temperatures), we showcase the ability of LLMs to predict material properties and discover materials with specified properties. 
Our results show that LLMs, even when trained on relatively small datasets, can outperform traditional ML approaches with carefully curated textual descriptions. Furthermore, the combination of domain-specific and general LLMs opens new avenues for materials discovery, while the incorporation of RAG enhances real-time learning and improves the scope and accuracy of LLM predictions.

Despite these promising results, several challenges remain, particularly in ensuring the stability of continuous quantitative predictions. 
For example, minor variations in temperature, such as between 500.12 K and 500.13 K, may lead to inconsistencies in property predictions like the bulk modulus. To address these issues, future work will focus on generating larger datasets~\cite{kaplan2020scalinglawsneurallanguage}, developing multi-agent systems for incremental task-solving~\cite{ghafarollahi2024atomagentsalloydesigndiscovery}, exploring novel digital encoding methods for LLMs~\cite{golkar2023xvalcontinuousnumberencoding}, and guiding LLMs to learn materials laws (such as the general trend of decreasing elastic constant tensor values with increasing temperature). 
These improvements, along with the addition of constraints or regularization techniques, may enhance the stability of numerical predictions.

Our work presents a fresh perspective on using LLMs for the quantitative prediction of material properties and facilitating inverse material design. 
A key benefit of domain-specific LLMs is the ability to interact with and generate results through natural language, without requiring users to have extensive knowledge of the underlying ML techniques. 
This lowers the barrier to entry for computational materials design and fosters broader participation in the field. 
The integration of domain-specific and general-purpose LLMs allows for access to broader research data, enhancing the synergy between materials science and AI. These advancements have the potential to revolutionize both fields by accelerating innovation, discovery, and application. 

\clearpage

\begin{figure*}[htbp]
  \centering
  \includegraphics[width=0.95\textwidth]{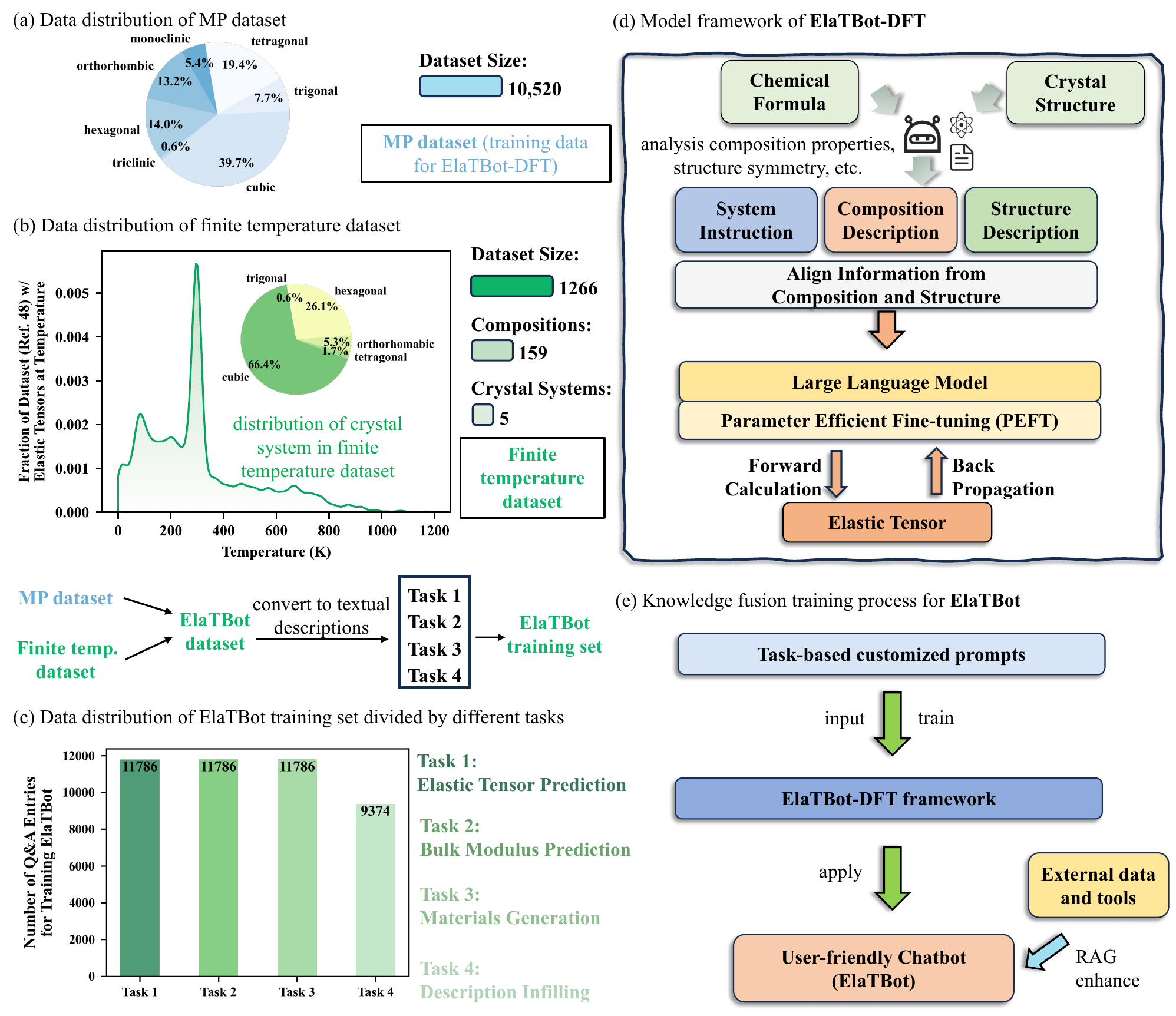}
  \caption{Dataset and model architecture for ElaTBot and ElaTBot-DFT. (a, b) Data distribution for Materials Project (MP) 0K DFT dataset and finite temperature dataset. For ElaTBot-DFT, we used data from the MP dataset, converting it to prompt type 4 (shown in Table S1). The transformed textual descriptions were separated to the training set, validation set, and test set and then used for training ElaTBot-DFT. For ElaTBot, as shown in the lower part of (b), we combined data from the MP dataset and the finite temperature dataset, then converted it into question and answer (Q\&A, each Q\&A pair is a task and we designed four tasks) as input and output for training ElaTBot, enabling ElaTBot to acquire multiple capabilities. (c) Number of Q\&A entries used for training ElatBot, categorized by the specific tasks in the training. The elastic constant tensor prediction task involves training the ElatBot to predict elastic constant tensors based on textual descriptions of materials. The bulk modulus prediction task requires the ElatBot to predict the bulk modulus from material textual descriptions. The material generation task aims to enable the ElatBot to generate material chemical formulas based on given bulk modulus and temperature. The description infilling task, given a description of the chemical formula and compositions, masks the formula with [MASK], and the ElatBot is then expected to fill in [MASK] with the correct chemical formula. (d) Model architecture for training ElaTBot-DFT. (e) Knowledge fusion training workflow for ElaTBot, detailing how external data and tools are integrated to enhance model capabilities.}\label{fig5}
\end{figure*}

\section*{Methods}
\subsection*{Data acquisition and processing}
We trained ElaTBot-DFT, Darwin, the random forest model and the MatTen model using material data containing elastic constant tensors from the Materials Project (MP). 
Initially, 12,128 materials with elastic constant tensor data calculated by DFT were available from MP (Fig.~\ref{fig1}(a)). 
After filtering out unreasonable entries, 10,520 valid samples remained. 
From this set, we allocated 9,498 to the training set, 500 to the validation set used during training, and 522 to the test set (521 for the random forest model due to the `NaN' problem in matminer). 
The distribution of materials with different crystal systems is shown in Fig.~\ref{fig5}(a). 
We compared the performance of all models on the test set using the mean absolute error (MAE) and the coefficient of determination ($R^2$). 
To prepare the data for model input, we followed  procedures appropriate for each model. 
For ElaTBot, we constructed textual descriptions based on the scheme in Table~S1, using pymatgen~\cite{ONG2013314} and robocrystallographer~\cite{Ganose2019} to convert composition and structural information into textual descriptions (see Fig.~S7). 
Darwin was trained with the same prompt type 4 used for ElaTBot. 
The random forest model was trained using Magpie feature vectors, which were derived from the elemental composition with pymatgen and matminer~\cite{WARD201860}. 
For the MatTen model, we constructed crystal structure graph neural networks following the original settings~\cite{D3DD00233K}.

In addition to the 10,520 data points for elastic constant tensors at 0K, we manually extracted 1,266 experimental elastic constant tensor data points at finite temperatures from~\cite{simmons1971single}. 
The distribution of elastic constant tensor data at different temperatures for this dataset is shown in Fig.~\ref{fig5}(b). 
To enable multitasking  in ElaTBot, we designed four tasks (Fig.~\ref{fig5}(c)) and converted material composition and structural information into textual descriptions as outlined in Table~S4. 
Given the limited availability of finite-temperature data, we did not create a separate test set for this subset. 
Instead, we evaluated predictive performance on unseen alloy compositions, including cubic phase $\text{Ni}_{3}\text{Al}$, $\gamma'-\text{PE16}$, and tetragonal phase $\gamma-\text{TiAl}$.

\subsection*{Model training and evaluation}
We trained the ELaTBot-DFT and ElaTBot models using an NVIDIA V100, both based on the Llama2-7b pre-trained model. 
The Darwin model, also built on Llama2-7b, was fine-tuned using a material science literature database. 
For comparison, we implemented the random forest model using scikit-learn~\cite{pedregosa2018scikitlearnmachinelearningpython} and employed a crystal structure graph neural network for the MatTen model. 
The training process for ElaTBot-DFT is shown in Fig.~\ref{fig5}(d), and that of ElaTBot is shown in Fig.~\ref{fig5}(e). 
To ensure a fair comparison, we standardized both the training duration and the number of samples processed across all models. 
Techniques such as early stopping were used to finalize the model when performance gains stagnated. 
Hyperparameters for model training are detailed in the Supplementary Materials (SM).

For fine-tuning, we applied LoRA+~\cite{hayou2024loraefficientlowrank}, a parameter-efficient adaptation technique that extends basic LoRA~\cite{hu2021loralowrankadaptationlarge}. 
LORA+ allows the adapter matrices to be fine-tuned at different learning rates, reducing GPU memory usage by approximately half without compromising data input capacity, thereby accelerating training. 
A detailed comparison between LoRA and LoRA+ is provided in Fig.~S8.

The LLMs (ElaTBot-DFT, ElaTBot, and Darwin) were managed using the Llama-factory~\cite{zheng2024llamafactory} framework, which facilitates model loading and parameter tuning. 
The random forest and MatTen models were trained directly using Python and PyTorch. 
The LLMs were optimized by calculating cross-entropy loss, $\mathcal{L}_{\text{CE}} = -\sum_{i=1}^{N} y_i \log(p_i)$, where $y_i$ is the true label and \(p_i\) is the predicted probability. 
The random forest model used squared error loss, $\mathcal{L}_{\text{SE}} = \frac{1}{N} \sum_{i=1}^{N} (y_i - \hat{y}_i)^2$, where \(y_i\) is the true value and \(\hat{y}_i\) is the predicted value. 
The MatTen model employed mean squared error loss $\mathcal{L}_{\text{MSE}} = \frac{1}{N} \sum_{i=1}^{N} (y_i - \hat{y}_i)^2$ for optimization, consistent with the method specified in~\cite{D3DD00233K}. 
These losses guide the models in learning to accurately predict the elastic constant tensor.

The predicted elastic constant tensor is expressed in Voigt form:
\[
\mathbf{C}_{ij} = \begin{bmatrix}
C_{11} & C_{12} & C_{13} & C_{14} & C_{15} & C_{16} \\
C_{21} & C_{22} & C_{23} & C_{24} & C_{25} & C_{26} \\
C_{31} & C_{32} & C_{33} & C_{34} & C_{35} & C_{36} \\
C_{41} & C_{42} & C_{43} & C_{44} & C_{45} & C_{46} \\
C_{51} & C_{52} & C_{53} & C_{54} & C_{55} & C_{56} \\
C_{61} & C_{62} & C_{63} & C_{64} & C_{65} & C_{66}
\end{bmatrix}
\]

The average value of elastic constant tensor $C_{ij}$ are calculated as follows:
\[
\overline{C_{ij}} = \frac{1}{36} \sum_{i=1}^{6} \sum_{j=1}^{6} C_{ij}
\]

The bulk modulus $K$ are calculated by pymatgen~\cite{ONG2013314} as follows:
\[
\text{K}_{voigt} = \frac{1}{9} \sum_{i=1}^{3} \sum_{j=1}^{3} C_{ij}
\]

The mean absolute error (MAE) and coefficient of determination ($R^2$) are used to evaluate the model performance. The MAE is calculated as follows:
\[
\text{MAE} = \frac{1}{n} \sum_{i=1}^{n} |y_{i} - \hat{y}_{i}|
\]
$R^2$ is 
\[
{R^2} = 1 - \frac{\sum_{i=1}^{n} (y_{i} - \hat{y}_{i})^2}{\sum_{i=1}^{n} (y_{i} - \bar{y})^2}
,\]
where  $y_{i}$ represents the labeled average value of the elastic constant tensor or bulk modulus from the dataset for each data point $i$. 
$\hat{y}_{i}$ is the predicted average value of the elastic constant tensor or bulk modulus from the models for each data point $i$. 
The symbol $\bar{y}$ represents  the mean of $y_{i}$ across all data points, and $n$ is the total number of data points.
The symmetry of the elastic constant tensor is checked by comparing the predicted tensors with the Voigt format matrix. 
Different crystal symmetries imply that certain components of  ${C}_{ij}$ are zero and specific relations exist between  ${C}_{ij}$ components (e.g., see 
~\cite{LI2022108280}).

For finite-temperature predictions, ElaTBot generated elastic constant tensors for \(\text{Ni}_{3}\text{Al}\) and \(\gamma^{'}-\text{PE16}\) at $T = $ 90, 113, 142, 162, 192, 223, 253, 283, 303, 333, and 363 K, and for \(\gamma-\text{TiAl}\) at $T = $ 30, 50, 70, 90, 110, 130, 150, 170, 190, 210, 230, 250, 270, 290, and 298 K. 
Due to the limited training data and the discrete nature of the prediction of ElaTBot, we did a linear fit to the predicted values and used this to evaluate the deviation from experimental data and analyze error trends.

\subsection*{Materials generation and RAG-enhanced prediction}
We used gradio~\cite{abid2019gradiohasslefreesharingtesting} to build a user-friendly chat interface for interacting with ElaTBot. 
This interface allows users to predict, generate, and perform RAG-enhanced prediction tasks through natural language input without the need to engage directly with code. 
While a default prompt, used during training, is pre-loaded into the interface,  users can modify it as needed for specific tasks. 

The RAG-enhanced prediction ability of ElaTBot was enabled through the integration of RAG, which allows the model to perform real-time learning without requiring retraining. 
The RAG module was implemented using langchain~\cite{pandya2023automatingcustomerserviceusing}, and follows a multi-step process including document loading, splitting, storage, retrieval, and output generation. 
This process enhances the ability of ElaTBot to update its knowledge and handle new data efficiently, as outlined in Fig.~\ref{fig3}(a).

\section*{Data availability}
The method section provides the models and algorithms employed in this study, while specific parameter implementations are available in Supplementary Materials.

\section*{Code availability}
All codes used in the paper will be publicly accessible on GitHub, and these codes will be provided upon reasonable requests.

\def\bibsection{\section*{\refname}}
\bibliography{./manuscript.bib}

\section*{Acknowledgments}
This work is supported by Research Grants Council, Hong Kong SAR through the General Research Fund (17210723,17200424). 
T.W. acknowledges additional support by The University of Hong Kong (HKU) via seed funds (2201100392, 2409100597).

\section*{Competing interests}
The authors declare no competing interests.

\end{document}